# Traffic Hotspot localization in 3G and 4G wireless networks using OMC metrics

Aymen Jaziri
Orange Labs,Telecom Sudparis
Email: aymen.jaziri@orange.com

Ridha Nasri
Orange Labs
Email: ridha.nasri@orange.com

Tijani Chahed
Telecom Sudparis
Email: tijani.chahed@telecom-sudparis.eu

*Abstract*—In recent years, there has been an increasing awareness to traffic localization techniques driven by the emergence of heterogeneous networks (HetNet) with small cells deployment and the green networks. The localization of hotspot data traffic with a very high accuracy is indeed of great interest to know where the small cells should be deployed and how can be managed for sleep mode concept. In this paper, we propose a new traffic localization technique based on the combination of different key performance indicators (KPI) extracted from the operation and maintenance center (OMC). The proposed localization algorithm is composed with five main steps; each one corresponds to the determination of traffic weight per area using only one KPI. These KPIs are Timing Advance (TA), Angle of Arrival (AoA), Neighbor cell level, the load of each cell and the Harmonic mean throughput (HMT) versus the Arithmetic mean throughput (AMT). The five KPIs are finally combined by a function taking as variables the values computed from the five steps. By mixing such KPIs, we show that it is possible to lessen significantly the errors of localization in a high precision attaining small cell dimensions.

## I. Introduction

One of the most important tasks that a telecommunication professional has is the network planning and optimization especially with the new technologies proposed in 3GPP standards like HetNet deployment [1] and the features of green networks[2]. RF planning and optimization mainly consist in the configuration and the densification of the network by modifying or adding new small cells and/or changing the network parameters and also saving energy by switching on/off [3] parts of the network according to the load in each zone. These tasks derive the need of finding hotspot zones [4] where the data traffic is very high leading thus to the congestion of the network and the degradation of its performances. Besides, if we don't localize these zones properly - like choosing wrong areas - the worst consequences range from an overall bad system performance (compared to what it could be) to cases where we need more equipment to meet the requirements in the same region. In other words, it implies in loss of CAPEX, OPEX and poor network quality.

Heavy traffic localization often consists in using probes. Measurements are performed to optimize the network configuration and deploy new small cells or switch on/off some cells. However, it is possible to identify the heavy traffic areas in the network using KPIs taken from the OMC. The accuracy of localization can be very high and improved comparing to the recent used techniques.

As a new alternative, key performance indicators can give an idea about the hotspot zones but with very low accuracy (with a granularity of the cell radius). Nevertheless, the combination of many KPIs would significantly increase the precision of traffic localization.

The main contribution of this paper is to combine several KPIs according to a given new algorithm. This new algorithm is not clearly disclosed because of the intellectual property. Using this new method, the precision of localization is increased and the costs are significantly reduced compared to probes-based methods.

Several traffic location techniques have been proposed to detect the hotspot zones of the data traffic. Currently, the operational methods used for optimization are based on probing, trace analysis and protocol decoding. Probing is an expensive solution because it needs tools for processing probes, storage servers... In the other hand, some solutions are under studies in the literature in order to give the highest accuracy with the lowest costs. To be based on the real data traffic evolution, authors in [5] provide a test transmitter which plays the role of a neighbor cell at first and then the role of a serving cell. Tests are realized within the existing cells in the network in order to assess the traffic density within the vicinity of the transmitter means. This solution needs to realize measurements in each area separately rendering it cumbersome and also comparable to probing in terms of cost. In the same context, the patent [6] discloses a method where the user equipments (UEs) send periodically a radio measurement report taken from both the serving cell and the neighbor cells. A recording unit is installed on the interface between the base station (BS) and studied base station controller (BSC) interface (A-BIS), to examine the messages exchanged on this interface, and based on these measurements, the traffic distribution is calculated. This method is not very accurate and the dimension of precision doesn't go to the small cell dimensions. In [7], authors propose to divide each cell of the network in a plurality of areas, and calculate a traffic value for each of these areas from the measurements in the network. The useful indicators to compute this traffic values are the channel occupation time, the call attempts, the number of handovers. This method is very simple since it is easy to get the inputs needed and it doesn't need additional tests or equipment. However, the localization of hotspots with this method is not very accurate and also the proposed optimization problem could be hard to solve. Another alternative is presented in the patent [8] which treats the hotspot traffic localization via statistical analysis of the timing advance measurements. Then, the accuracy is increased with providing the related neighbor cell measurements. These measurements are collected and processed in a base station controller BSC and the related accuracy is better than the previous solutions but still not enough to meet with dimensions of small cells. With probing like in [5] [6] [8], the costs are important due to the equipment used to collect information from calls and the need for servers to store the traces. Also, the probes could not collect all the traffic transmissions, some of the transmissions are not captured and lost, so it is still not too efficient for traffic localization. However, a solution based on KPIs like in [7] will not cost too much; we have only to provide an algorithm to localize the heavy traffic with high accuracy.

Note that the localization of each UE is possible and with a good accuracy thanks to many techniques cited in [9]. For example, triangulation with AoA measurement (LTE Rel 9) is used and means that the AoA of the UE signal is calculated from two reference points which are generally the position of two BSs then the location is determined by the intersection of the two lines bearing from the known reference points. However, these individual locating methods are cumbersome and consume a lot of time to extract from them the distribution of traffic data transmission because it deals with a large number of users in the network.

This paper is organized as follows. First, the KPIs, from which we extract the information of traffic distribution in the network, are defined. Next, a modular algorithm is proposed and in each step, only one KPI is analyzed before combining them in the last step. Then, the performance of the estimation is analyzed by simulating an LTE network with the configuration recommended in 3GPP-LTE A standard.

## II. Description and design of the traffic localization algorithm

### A. Network used metrics

In order to better understand the utility of all the aforementioned KPIs, each KPI is calculated and can be obtained a follows:

*1) Timing Advance:* For the timing advance (in GSM, LTE and LTE-A) or the propagation delay (in WCDMA), a time offset measurement is realized by the BS between its own transmission and the transmission received from the UE. According to the calculated offset, the serving BS determines the suitable timing advance for the UE [10]. Then, from this timing advance, the BS calculates the distance traveled by the radio signal. In practice, depending on the resolution (or granularity) of timing advance, a specific distance range where the UE is located will be calculated. Actually, timing advance is used as a KPI for network supervision and analysis and with a granularity of 78.125 meters in LTE networks [10]. For the timing advance, each cell is divided into intervals according to the interval of distances like in figure 1. Then the distribution of the UEs in the cell is calculated according to these intervals. For example, 30% of the UEs are in the range of TA1, 20% in the range of TA2, 40% in the range of TA3 and 10% in the range of TA4.

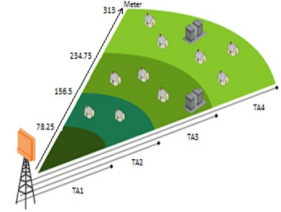

Fig. 1. The division of the covered area according to TA

*2) Angle of Arrival:* Angle of Arrival (AoA) is defined as the estimated angle of a UE with respect to a reference direction which is defined as geographical North; the value of AoA is positive in an anticlockwise direction [11]. In general, any uplink signal from the MS can be used to estimate the AoA, but typically a known signal such as the Sounding Reference Signals (SRSs) or DeModulation Reference Signals (DM-RS) would be used [12]. In order to construct the shape of the downlink beam, direction of arrival estimation already exists and is supported in WCDMA but it is not standardized [13].

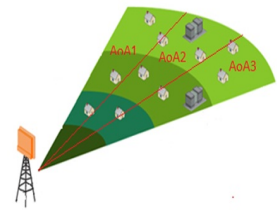

Fig. 2. The division of the covered area according to AoA

According to the angle of arrival, the cell is divided into ranges like in figure 2 relative to the angle between the UE, the BS and the geographical north. With this classification, the corresponding spatial distribution is calculated. An example of distribution that we can have is 30% of UEs with range corresponding to AoA1, 40% for AoA2 and 30% for AoA3.

*3) Neighboring Cell Level:* In each access or handover process, every UE measures the signals of the detected cells and sends a report of these measurements to its serving cell. In GSM, the detected cells list is reported periodically by the UE to the BS. However, in 3G and 4G networks, it is reported to the BS only when a special event is activated like handovers or cell selection. For each neighboring cell, a percentage of UEs, connected to a serving cell and declare this neighboring cell as the first best candidate for handover, is calculated. The corresponding KPI is the percentage of UEs declaring the same neighbor cell as their second best serving cell (best candidate cell for handover). The information extracted from this KPI is exploited to recover the errors of AoA estimation which is due to reflexions in real networks. The exploitation of this KPI is equivalent to the triangulation technique used in the probes-based methods.

*4) Load Time:* The load time represents the percentage of time when the serving cell's resources are fully occupied. This KPI is a gauge counter (means it is a percentage in time) already defined in the actual tools and is calculated during busy hours. In hotspot zones, the traffic evolution almost has the same behavior in the neighbor cells due to the load balancing and forced handovers [14]. Then, if a cell is congested and one of its neighbors is congested, it is most probable that most of the traffic is located in the edge between the two cells and if there is no correlation with any of the neighbor cells, the traffic is most probably generated from users close to the serving cell.

*5) Harmonic and Arithmetic Mean Throughput :* The harmonic mean throughput gives more importance to the UEs with bad radio conditions and which are generally in the cell edge. From the formula of the cell throughput defined into the OMC by some constructors, it is important to notice that the cell throughput has approximately the same value as the harmonic mean throughput. On the other side, the arithmetic mean throughput is the mean of the throughputs of all the connected users. The utility of these KPIs comes from the following possible scenarios; if most of the UEs are close to the BS (so, in very good radio conditions), the arithmetic mean throughput is high and the harmonic mean throughput is low. In this case, the difference between the harmonic and the arithmetic mean throughput will be high. However, when most of the UEs are in the cell edge, the arithmetic mean throughput decreases and the harmonic mean throughput increases comparing to the first scenario. So, the difference between the two mean throughput is low. In sum, the difference between the arithmetic and the harmonic mean throughput gives an idea about the location of the hotspot of traffic. This KPI is expected to give the same information like the correlation of load time but with different parameters and this can ameliorate more the estimation and correct the errors due to time dispersion or reflexions problems.

*B. Algorithm formulation*

Estimating the hotspot zones in the network is realized as follows: First, the OMC reports the statistics of each KPI corresponding to the cells in its controlled area. Then, each KPI is analyzed into an algorithm in order to extract weights indicating the spatial distribution of the traffic.

*Step 1: calculate the spatial distribution of weights according to the timing advance:*
Actually, the cell is divided into many areas depending on the distance (or the timing advance) between the pixel's position and the serving BS's position. Then, each pixel takes as a weight the percentage of UEs in its range of TA.

*Step 2: calculate the spatial distribution of weights according to the Angle of arrival:*
Each cell divides its covered area into many sub-areas depending on the angle between each pixel's position, the serving BS's position and the geographical north. So, each pixel takes as a weight the percentage of UEs in the same range of angle of arrival defined in the corresponding KPI.

*Step 3: calculate the distribution of weights according to the Detected neighbor cell:*
According to the neighboring cell level, the weight in each pixel will be the percentage of UEs reporting as a second best serving cell the same one as the cell having the best RSRP measurement in this pixel. It is indeed possible to define an RSRP map, and define the second best serving cell for each pixel with the RF fingerprinting technique [15]. In this context, RF fingerprinting consists in matching each pixel to a range of statistics pertaining to RF topology. With the grid calculated by this technique, second best serving cell is defined in each pixel.

*Step 4: calculate the distribution of weights according to the traffic load time and correlation between cells:*
The correlation in load time between cells is tested and the distribution of weights is calculated. In the following step, for each pixel, we identify the set of cells which are candidate to be the serving cell and with a load time near to the load time of the best serving cell. Then each pixel in the network map takes as a weight the sum of all the load time of the set divided by the cardinality of this set.

*Step 5: calculate the distribution of weights according to the Arithmetic mean throughput vs Harmonic mean throughput:*
The weights are the difference between the arithmetic mean throughput of the cell and the harmonic mean throughput. This quantity is divided by a constant in order to have all the calculated weights in $[0, 1]$. The weight in each pixel depends on its position relative to the position of its best serving cell and also on the value of the difference between the two mean throughput. So, the pixels in the cell edge take the value of the difference between the HMT and the AMT divided by a constant if this difference is less than a certain threshold which should be calculated properly and they take zero if this condition is not satisfied. In contrast, the weights in the pixels close to the BS take the value of the difference between the HMT and the AMT divided by a constant if this difference is higher than a certain threshold and zero else.

*Step 6: include all KPIs in a metric and estimate the hotspot traffic zones:*
After analyzing all the KPIs and extracting values from them, the most important step is how to transpose all the outputs from the previous steps of the main algorithm into the same map. In fact, a metric including all the KPIs is defined with the different weights depending on the importance of the information given by each KPI. The metric used can be any function including all these projections and it is possible to give more importance to some KPIs and to reduce the impact of the other KPIs.

Finding the best function can be an optimization problem that gives the optimal coefficients for the function.

In order to improve the estimation of the traffic distribution in the network, the values estimated in each pixel are combined with the values calculated in the neighboring pixels. The function of the smoother is distance decay because the correlation between two pixels is smaller when the distance between them is high.

### III. SIMULATION AND ANALYSIS

#### A. Simulation

In order to test the proposed algorithm, we consider a realistic network with 23 tri-sector sites (see figure 3) with the available bandwidth of 20MHz. At each second, UEs are generated depending on the arrival rate. We suppose that each UE has a file of size 1000 Kbit to download. Moreover, a part of UEs are moving during the transmission with mobility of 8.33 km/h meaning that the simulation supports also events of handover. UEs are scheduled according to the round robin model. Finally, KPIs are updated in the network and are filtered approximately every quarter of an hour which corresponds to the time granularity of most OMC vendors.

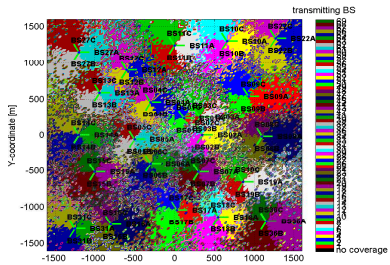

Fig. 3. Map of the network layout and the best serving cell in each pixel

#### B. Numerical results

In figure 4, we plot the real distribution of the traffic. In figure 5, the estimated traffic distribution is represented. Thus, the first conclusion is that the hotspot zones are found in the entire network map and the accuracy of the estimation is 25 meters.

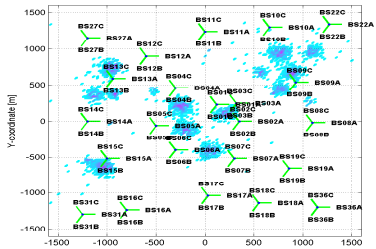

Fig. 4. Real data distribution

The figure 6 shows that using all the five KPIs in a real network topology gives better estimation of traffic distribution comparing to the case when using some of them. Moreover, from the figure 6, the TA, AoA and the neighbor cell level are useful KPIs but each one is with low accuracy, however the Mean throughput and the correlation of the load time are less important but they have also an impact on the estimation and they improve significantly the estimation when using the other KPIs and it doesn't improve the estimation when using some of these KPIs.

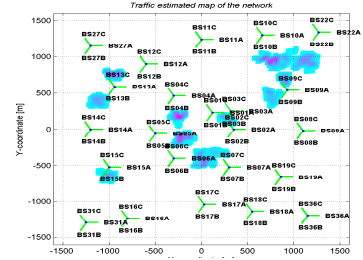

Fig. 5. Estimated data distribution

The error of estimation using all the proposed KPIs is less than 10% while in the case where one of the KPIs is eliminated; the error becomes more than 12%.

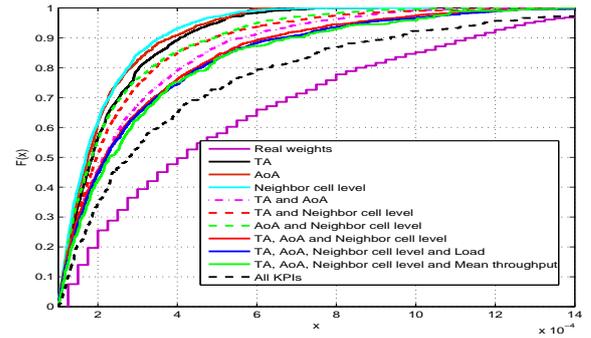

Fig. 6. CDF of real and estimated weights in term of access demand

For a moderate traffic, the performance of our estimation can be observed in the figure 6. However, there stills some problems related to these KPIs especially when the data traffic is heavy. So, some of the KPIs can be biased by the UEs not accepted in the network and trying again and again to be served. This problem can be treated with analyzing only the performance in term of elapsed traffic. Then, the figure 7 represents the CDF of elapsed traffic during the simulation. As a result, simulations proved that exploiting all the proposed KPIs gives the smallest error of less than 8% in the estimation of elapsed traffic and the percentage of error is less than the case where one of these KPIs is eliminated.

We compare in the following table the percentage of detected hotspots which has the highest weights to the real hotspots in the network.

From the previous table, we observe that the first hotspots which have the highest weights are well estimated. In other words, the weights attributed to 0.5% of most important hotspot zones represent 0.42% hotspot zones in the estimation. However, when the threshold of detection is decreased, the estimation presents less efficiency and this is due to the fact that some zones near to the hotspots could take a significant weight without carrying really heavy traffic. In the other side, the results of the proposed algorithm are better than in the case where only TA is exploited which is a solution proposed in several studies [8].

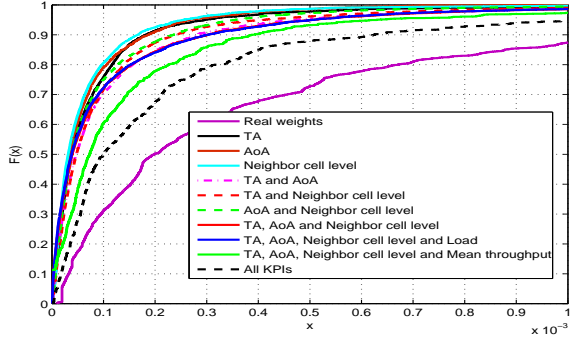

Fig. 7. CDF of real and estimated weights in term of elapsed traffic

Table: Percentage of detected hotspots

| Real % of first hotspots | Estimated % with all KPIs | Estimated % with only TA |
|---|---|---|
| 0.5% | 0.42% | 0.12% |
| 1% | 0.57% | 0.26% |
| 2% | 1.1% | 0.47% |
| 5% | 2.51% | 1.12% |
| 10% | 4.74% | 2.29% |
| 20% | 9.55% | 4.87% |
| 50% | 26.92% | 14.22% |
| 70% | 41.36% | 23.23% |

In sum, the simulations show encouraging results and the proposed solution is very promising to meet with the exact results with optimizing the function including all the KPIs and 10% of error in the hotspot zones is near to the percentage of error generated by the actual used techniques such as probing which costs too much in term of OPEX and CAPEX.

## IV. Conclusion

Traffic hotspot localization is an important issue in the network densification and also in the efficient policy of switching on/off small cells. Therefore, proposing a new technique based only on OMC metrics could be a very interesting solution that can save more CAPEX and OPEX and give a good precision. In this work, a new algorithm consisting in exploiting KPIs is built and the accuracy that it provides is high comparing to the actual and operational tools. It can be considered as a new tool for dimensioning and planning small cells. KPIs' based location solution shows a good precision with a good level of certainty.

For future works, we aim to find the suitable function including the KPIs and this can be done by formulating an optimization problem which gives the best function mixing these KPIs with the optimal weights. Furthermore, potential hotspots and additional information layer about commercial or industrial zones would be treated in order to optimize much more the proposed localization technique.